\documentclass[aps,prb,twocolumn,superscriptaddress]{revtex4}

\input epsf.sty

\begin{document}


%
%

\title{Neutron scattering study of soft phonons and diffuse scattering in insulating
La$_{1.95}$Sr$_{0.05}$CuO$_{4}$}

\author{Shuichi Wakimoto}
\email[Corresponding author: ]{waki@physics.utoronto.ca}
\affiliation{ Department of Physics, University of Toronto, Toronto,
   Ontario, Canada M5S~1A7 }

\author{Seunghun Lee}
\affiliation{ NIST Center for Neutron Research, National Institute
of Standards and Technology, Gaithersburg, Maryland 20889-8562, USA}

\author{Peter M. Gehring}
\affiliation{ NIST Center for Neutron Research, National Institute
of Standards and Technology, Gaithersburg, Maryland 20889-8562, USA}

\author{Robert J. Birgeneau}
\affiliation{ Department of Physics, University of Toronto, Toronto,
   Ontario, Canada M5S~1A7 }

\author{Gen Shirane}
\affiliation{ Department of Physics, Brookhaven National Laboratory,
   Upton, New York 11973-5000, USA }

\date{\today}

\begin{abstract}

Soft phonons and diffuse scattering in insulating
La$_{2-x}$Sr$_{x}$CuO$_4$ ($x=0.05$) have been studied by the
neutron scattering technique.  The X-point phonon softens from high
temperature towards the structural transition temperature
$T_{s}=410$~K, and the Z-point phonon softens again below 200~K. The
Z-point phonon softening persists to low temperature, in contrast to
the behavior observed in the superconducting $x=0.15$ compound, in
which the Z-point phonon hardens below $T_c$. The diffuse scattering
associated with the structural phase transition at 410~K appears at
commensurate positions.  These results highlight interesting
differences between superconducting and insulating samples.

\end{abstract}


\maketitle

\section{Introduction}

No direct connection between the unusual superconductivity and the
lattice dynamics in the cuprate high-$T_c$ materials has been
established to date.  By contrast, two of the most successful
demonstrations of such a connection in BCS materials were reported
for Nb$_3$Sn (Ref.~\onlinecite{Axe_73}) and 
for Nb (Ref.~\onlinecite{Shapiro_75}) by 
neutron inelastic scattering techniques to show that acoustic
phonons with energies less than the superconducting gap energy
soften and exhibit linewidths that decrease below $T_c$. The smaller
linewidths correspond to longer phonon lifetimes, and result from
the absence of any decay channel available to such phonons in the
presence of the superconducting gap which opens up at $T_c$.
Although such behaviour has not been confirmed for the high-$T_c$
cuprate La$_{2-x}$Sr$_{x}$CuO$_4$ (LSCO) system, low-energy soft
phonons have been reported to show interesting correlations with the
superconductivity.  Firstly, an anomalous lattice hardening below
$T_c$ has been observed by ultrasonic measurements~\cite{Nohara_93}.
Later, it was found that the soft phonon associated with the tilting
mode of the CuO$_6$ octahedra hardens, and that the phonon linewidth
saturates below $T_c$.~\cite{CHLee_96,Kimura_00}

Another interesting feature has been studied in the
high-temperature-tetragonal (HTT) phase of LSCO with $x=0.12$, where
Kimura {\it et al.}~\cite{Kimura_00} have found that the diffuse
scattering appears at incommensurate (IC) lattice positions located
around the X-points, $(n/2, n/2, 2l)$, as shown by the open circles
in Fig. 1 (a).  Intriguingly, the incommensurability saturates at
the same value as that exhibited by the IC static (or dynamic)
antiferromagnetic peaks, which suggests that a correlation may exist
between the superconductivity, magnetism, and lattice distortion.
In this paper, we report the results of neutron scattering
experiments performed on a single crystal sample of insulating
La$_{1.95}$Sr$_{0.05}$CuO$_4$ that were designed to test if the
features of the soft phonons and the diffuse scattering mentioned
above are in fact related to the superconductivity.  Our results
bring to light an interesting contrast between superconducting and
insulating samples of LSCO.

\begin{figure}
\centerline{\epsfxsize=3.4in\epsfbox{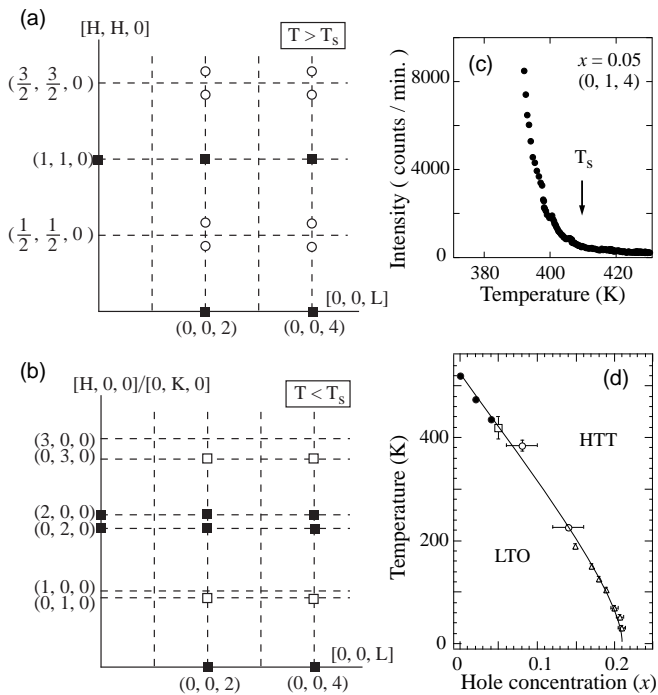}} 
\caption{The
reciprocal lattice of LSCO is shown at temperatures (a) above and
(b) below the structural transition temperature $T_{s}$. Closed
squares represent nuclear Bragg peaks, while open squares represent
superlattice peaks in the LTO phase. The open circles correspond to
the IC diffuse scattering in the HTT phase observed for $x=0.12$ by
Kimura {\it et al}.~\cite{Kimura_00} (c) Temperature dependence of
the superlattice peak intensity for $x=0.05$.  (d) $T_{s}$ as a
function of doping $x$. The lone open square corresponds to the
crystal studied in this paper. The closed circles, open circles, and
open triangles represent data taken from
Ref.~\onlinecite{Keimer_92}, \onlinecite{Thurston_89}, and
\onlinecite{Takagi_92}, respectively.}
\end{figure}

\section{Scattering geometry and experimental details}

The high-temperature structure of LSCO is tetragonal (space group
$I4/mmm$), and it is known that an instability of the zone boundary
X-point soft phonon causes a displacive structural transition to the
low-temperature orthorhombic (LTO) phase (space group
$Bmab$).~\cite{Birgeneau_87,Boni_88,Thurston_89}  The left side of
Fig. 1 depicts the reciprocal lattice of the HTT phase ($T>T_{s}$,
where $T_{s}$ is structural transition temperature), and the LTO
phase ($T<T_{s}$).  Due to the twinned structure of the crystal, the
$[HH0]$ axis in the HTT phase becomes a superposition of the $[H00]$
and $[0K0]$ axes in the LTO phase.  In the HTT phase, IC diffuse
scattering has been observed for a superconducting $x=0.12$ sample
around the X-points as shown by open circles.~\cite{Kimura_00}  In
the LTO phase, superlattice peaks appear at reciprocal lattice
points such as (014) and (032), shown by the open squares, which
become zone-center $\Gamma$-points, while positions such as (104)
and (302) become zone-boundary Z-points.

Thus the structural transition lifts the degeneracy of the X-point
phonon, and two modes can be distinguished in the LTO phase, i.e.
the $\Gamma$-point and the Z-point phonons, which correspond to the
octahedral tilting mode along the orthorhombic $b$- and $a$-axis,
respectively. The $\Gamma$- and Z-point phonons harden upon cooling
below $T_{s}$, but then the Z-point phonon softens subsequently at lower
temperatures, implying an instability against a low-temperature
tetragonal ($P4_2/ncm$) or low-temperature less-orthorhombic
($Pccn$) phase.~\cite{Thurston_89}  In superconducting samples it
has been reported that this Z-point softening breaks at $T_c$,
suggesting that the superconducting state stabilizes the LTO
phase.~\cite{CHLee_96,Kimura_00}

In the present study, Z-point phonons were measured at the (104) and
(302) reciprocal lattice points, respectively, using the SPINS and
BT9 spectrometers located at the NIST Center for Neutron Research.
The (104) Z-point phonon was measured using a fixed final neutron
energy $E_f=5$~meV and horizontal beam collimations
32$'$-80$'$-S-80$'$-open (S = sample), while the (302) Z-point
phonon was measured using a fixed initial neutron energy
$E_i=14.7$~meV and collimations of 40$'$-22$'$-S-24$'$-open. In
addition, X-point phonons were studied at (1/2,1/2,4) (tetragonal
indexing) using SPINS with the same instrument configuration except
with a horizontally-focusing analyzer in place of the flat analyzer.
A Be filter placed after the sample, and a PG filter placed before
the sample, were used to eliminate higher order ($\lambda/2$,
$\lambda/3$, ...) neutrons for the $E_f=5$~meV and $E_i=14.7$~meV
measurements, respectively.

The single crystal of La$_{1.95}$Sr$_{0.05}$CuO$_4$ used for the
present study was grown in a floating zone
furnace,~\cite{Hosoya94,CHLee_98} and is the same sample that was
used for the neutron scattering measurements of the diagonal IC
magnetic peaks.~\cite{waki_rapid,waki_full} Lattice constants of the
LTO phase at room temperature are $a_o=5.38$~\AA~ and
$c_o=13.24$~\AA, which correspond to reciprocal lattice units of
$a_o^*=1.17$~\AA$^{-1}$ and $c_o^*=0.47$~\AA$^{-1}$.  The lattice
constant in the HTT phase is related to that of the LTO phase by
$a_t \sim a_o/\sqrt{2} = 3.80$~\AA, thus $a_t^* \sim
1.65$~\AA$^{-1}$. The structural transition temperature $T_{s} \sim
410$~K was determined by monitoring the temperature dependence of
the superlattice peak intensity at (014), the data for which are
shown in Fig. 1 (c). This value is in excellent
agreement with the $x$-dependence of $T_{s}$ accumulated from
previous measurements~\cite{Thurston_89,Keimer_92,Takagi_92} shown
in Fig. 1 (d), where the $x=0.05$ data point is
represented by an open square.

\begin{figure}
\centerline{\epsfxsize=3.5in\epsfbox{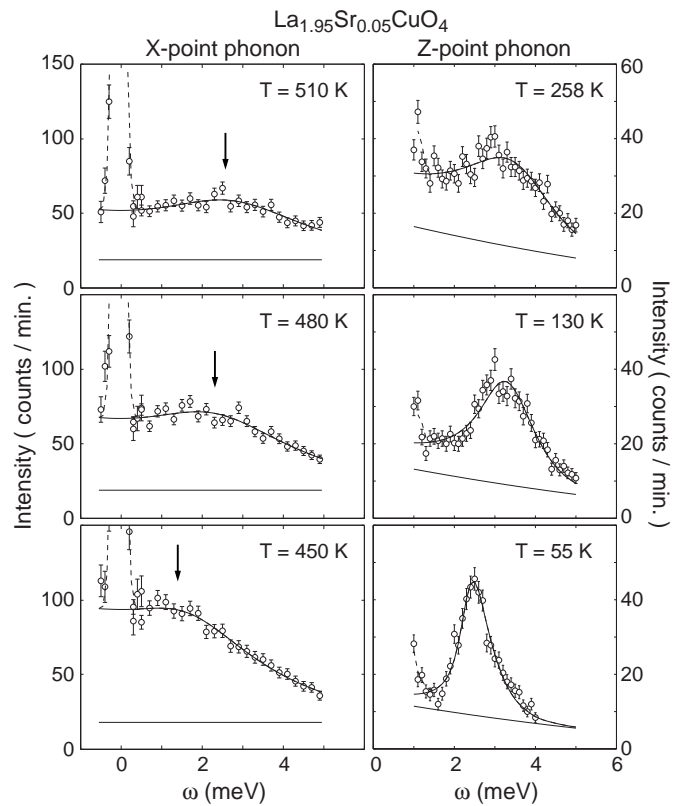}}
\caption{Left-hand panels show X-point phonons measured at ${\rm \bf
Q}=(1/2, 1/2, 4)_{tetra}$ at three temperatures using the SPINS
spectrometer with the final neutron energy fixed at 5~meV.
Right-hand panels show Z-point phonons measured at ${\rm \bf Q}=(3,
0, 2)_{ortho}$ at three temperatures using the BT9 spectrometer with
the initial neutron energy fixed at 14.7~meV. The solid lines are the 
results of fits
to the Lorentzian function specified in the text, convoluted with the
instrumental resolution function.}
\end{figure}

\section{Soft phonon and diffuse scattering}

Representative spectra of the X- and Z-point phonons are shown in
Fig. 2. While the X-point phonon for $T>T_{s}$ is extremely broad in
energy, it is nevertheless apparent that the phonon energy decreases
on cooling towards $T_{s}$.  Below $T_{s}$, the Z-point phonon is
well-defined, and its energy and linewidth decrease with decreasing
temperature.

Profiles taken at several temperatures were fit to an instrumental
resolution-convoluted $S({\rm\bf q}, \omega)$ function,
\begin{equation}
S({\rm \bf q}, \omega) = (n(\omega) + 1) I_{ph} L({\rm \bf q},
\omega) + I_G \exp(- \omega^2 / 2 \sigma^2) + I_{BG}
\end{equation}
\begin{equation}
L({\rm \bf q}, \omega) = \frac{\gamma^2}{(\omega - \omega_{ph})^2 +
\gamma^2} + \frac{\gamma^2}{(\omega + \omega_{ph})^2+\gamma^2},
\end{equation}
where $n(\omega)=(e^{\omega/k_{B}T}-1)^{-1}$, $\gamma$ represents
the phonon energy half-width at half-maximum.  The first term in Eq.
(1) represents the phonon cross-section, while the second Gaussian
term describes elastic peak at $\omega = 0$.  From the dispersion
relation, the $q$-dependent phonon energy $\omega_{ph}$ is given by
\begin{equation}
\omega_{ph}^2 = \omega_0^2 + a_xq_x^2 + a_yq_y^2 + a_zq_z^2,
\end{equation}
where $\omega_0$ is the phonon energy at either the X- or Z-point.
The dispersion constants $a_x=620$, $a_y=250$, and
$a_y=1130$~meV\AA~ are taken from Ref.~\onlinecite{Thurston_89}.
Fits of the data to the above function are shown in Fig. 2.  The
solid lines reflect the Lorentzian component of the soft phonon as
well as the background, while the dashed lines represent the $\omega
= 0$ Gaussians.  Note that the background for the Z-point profiles
slopes as a result of the analyzer correction factor $k_f^3/\tan
\theta_A$ ($k_f$ is final neutron wave vector and $\theta_A$ is the
analyzer angle with respect to the scattered beam), which must be
applied to fixed-$E_i$ scans.

\begin{figure}
\centerline{\epsfxsize=3.4in\epsfbox{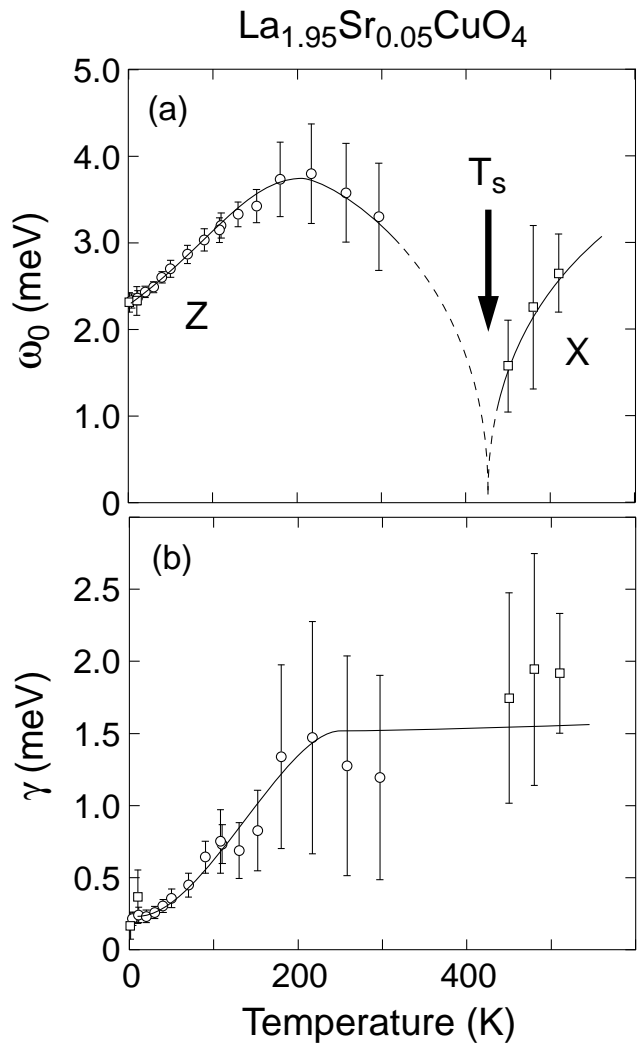}}
\caption{Temperature dependence of (a) the soft phonon energy
$\omega_0$ and (b) the phonon energy linewidth $\gamma$.  
The lines are guides-to-the-eye.  The squares and circles depict the
data taken at (1, 0, 4) (SPINS) and (3, 0, 2) (BT9), respectively.}
\end{figure}

The parameters obtained from the fits are summarized in Figs. 3 and 4. The
temperature dependence of $\omega_0$ and $\gamma$ are shown in Fig.
3 (a) and (b), respectively.  A general trend is observed on cooling
in which the X-point phonon softens towards $T_{s}$, and the Z-point
phonon recovers below $T_{s}$ and then softens again below $\sim
200$~K. Meanwhile, the linewidth $\gamma$ also starts to decrease
with decreasing temperature below $\sim 200$~K.  Around $T_{s}$, the
phonon profiles measured at SPINS (not shown) are not well-defined,
possibly because they overlap with a $\Gamma$-point phonon that may
not be high enough in energy to resolve from the Z-point phonon, and
in part because of the strongly damped nature of the soft phonons at
high temperatures.  Thus we were not able to determine if the phonon
softens completely at the structural transition temperature.

\begin{figure}
\centerline{\epsfxsize=3.5in\epsfbox{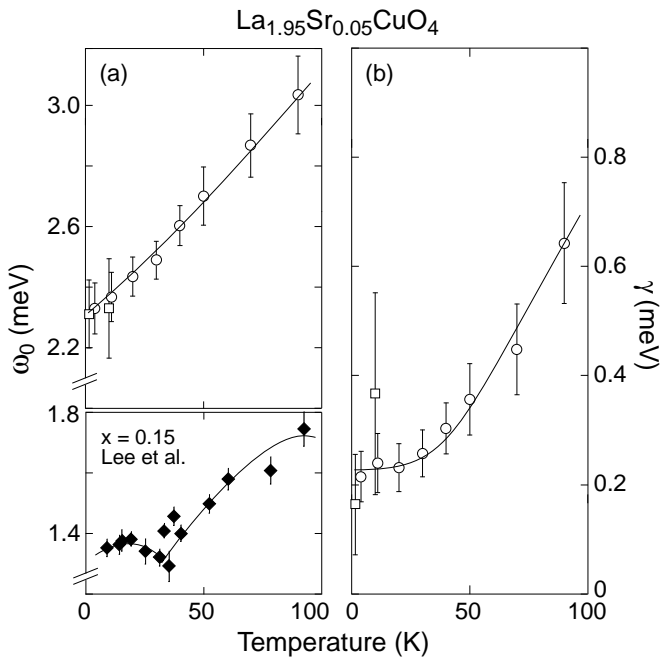}}
\caption{Enlarged figures of the same data in Fig. 3, covering 
the range $0 \leq T \leq 100$~K,
are shown.  Data for $x=0.15$ taken from
Ref.~\onlinecite{CHLee_96} are represented by diamonds in (a).
Lines are guides to the eye.  The squares and circles depict the
data taken at (1, 0, 4) (SPINS) and (3, 0, 2) (BT9), respectively.}
\end{figure}

An enlarged figure of $\omega_0$ that covers the temperature range
$0 \leq T \leq 100$~K is shown in Fig. 4 (a) together with the data
for $x=0.15$ reported by Lee {\it et al.}~\cite{CHLee_96}  The
$\omega_0$ of the insulating $x=0.05$ sample decreases monotonically
with decreasing temperature while the superconducting $x=0.15$
compound exhibits a softening that breaks at $T_c$.  On the other
hand, as shown in the enlarged Fig. 4 (b), the linewidth $\gamma$
for $x=0.05$ decreases with temperature and saturates below $\sim
30$~K, which is similar to the behavior of $\gamma$ in
superconducting samples, which saturates around $T_c \sim 35$~K
(Ref.~\onlinecite{CHLee_96,Kimura_00}). Thus soft phonons show both
contrasting and similar behavior in insulating and superconducting
samples.

\begin{figure}
\centerline{\epsfxsize=3in\epsfbox{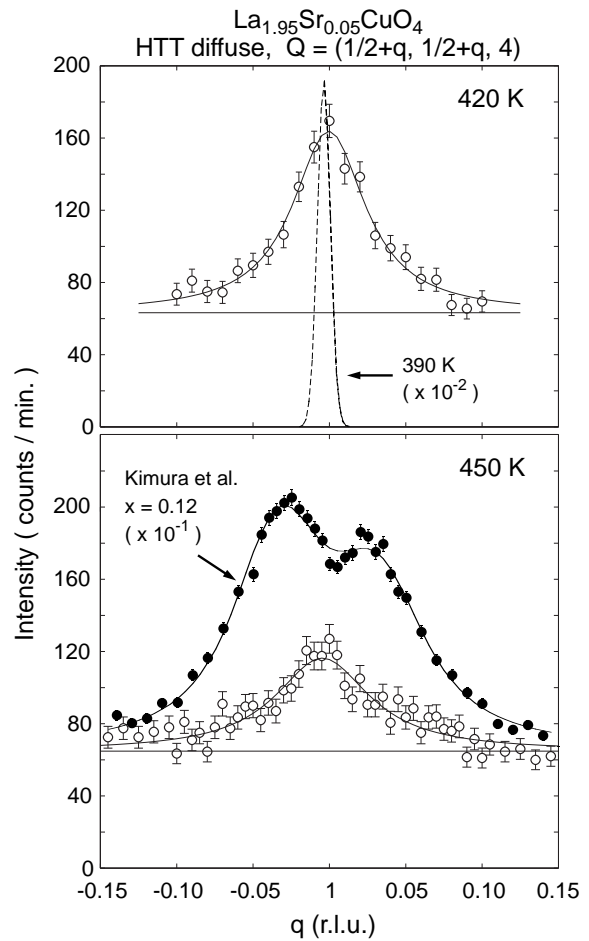}}
\caption{Diffuse
scattering profiles measured along the $(1/2+q, 1/2+q, 4)$
trajectory.  The superlattice peak measured at 390~K is also shown
as a dashed line in the upper figure.  The solid lines are fits to
the resolution-convoluted two-dimensional Lorentzian function plus
background.  The lower figure also shows IC diffuse scattering in
the HTT phase of the $x=0.12$ sample~\cite{Kimura_00}, which has a
$T_{s}$ of $240$~K.}
\end{figure}

Another important contrast between insulating and superconducting
materials is found in the diffuse scattering in the HTT phase.  As
mentioned in Sec. I, Kimura {\it et al.}~\cite{Kimura_00} have
reported IC diffuse scattering located around the X-point in the HTT
phase for a superconducting $x=0.12$ sample. However, we observe
commensurate diffuse scattering for an insulating $x=0.05$ sample.
Figure 5 shows diffuse scattering peaks along the $(1/2+q, 1/2+q,
4)$ direction for $x=0.05$ represented by open symbols. Commensurate
diffuse scattering grows with decreasing temperature towards
$T_{s}$, below which it is replaced with a sharp resolution-limited
superlattice peak shown by the dashed line in the upper figure.  The
lower figure also shows the IC diffuse peaks reported by Kimura {\it
et al.} for $x=0.12$ measured at 390~K along the $(3/2+q, 3/2+q, 2)$
trajectory for comparison.
The $x=0.05$ data are well-described by fits to a
resolution-convoluted two-dimensional Lorentzian centered at the
commensurate position, as shown by the solid lines with a half-width
at half-maximum $\kappa = 0.07$~\AA$^{-1}$ for 420~K and $\kappa =
0.09$~\AA$^{-1}$ for 450~K.

\section{Discussion}

We report soft phonon and diffuse scattering measurements on an
insulating $x=0.05$ LSCO sample with the goal of elucidating any
intrinsic relationship that may exist between the superconductivity
and lattice instability exhibited by these materials.  A comparison
between the data taken on the present insulating sample and
previously reported superconducting samples of LSCO reveals some
important contrasts.

Firstly, there is no break in the softening of the Z-point phonon in
the insulating sample (Fig. 4 (a)).  This reveals a robust
correlation between superconductivity and the structural instability
as discussed previously.~\cite{Nohara_93,CHLee_96,Kimura_00}  On the
other hand, the linewidths of the Z-point phonon for both insulating
($x=0.05$) and superconducting ($x=0.10$, 0.12, and 0.15
(Refs.~\onlinecite{CHLee_96,Kimura_00})) samples saturate at low
temperatures.  This suggests that the saturation of the linewidth is
not due purely to the superconducting state.  A possible alternative
explanation is that the saturated value is determined by impurity
scattering from the dopant Sr atoms, since the saturated value
increases monotonically with $x$.  The hole concentration $x$ and
saturated value of $\gamma_0$ are summarized in Table 1.  The
dependence of $\gamma_0$ is roughly linear with $x$.

Briefly, it might be also interesting to consider if there is any
significance to the maximum value of $\omega_0$ of the Z-point
phonon. Needless to say, the temperature where this maximum value is
achieved changes with the transition temperature $T_{s}$.  However,
for both $x=0.05$ and $0.15$ samples, the Z-point phonon starts to
soften at the temperature where the linewidth $\gamma$
coincidentally starts to decrease, suggesting a general correlation
between phonon lifetime and lattice instability.

\begin{table}
  \caption{Hole concentration $x$ and saturated linewidth $\gamma_0$.
  Note that the linewidth in Ref.~\onlinecite{CHLee_96} is defined as
  $2\gamma$.}
\begin{ruledtabular}
\begin{tabular}{lcc}
$x$
&
$\gamma_0$~(meV)
&
Ref.   \\
\hline
$0.0$
&
$0.10$
&
\onlinecite{CHLee_96} \\
$0.05$
&
$0.22$
&
present \\
$0.10$
&
$0.38$
&
\onlinecite{Kimura_00} \\
$0.12$
&
$0.55$
&
\onlinecite{Kimura_00} \\
$0.15$
&
$0.38$
&
\onlinecite{CHLee_96} \\
\end{tabular}
\end{ruledtabular}
\end{table}

Secondly, the diffuse scattering in the HTT phase is commensurate in
the insulating sample.  This suggests that the incommensurate nature
of the diffuse scattering is characteristic of the superconducting
sample, or characteristic near the specific concentration $x=0.12$
where the IC spin density wave order is well established at low
temperatures.~\cite{Suzuki_98,Kimura_99}  Moreover, the
incommensurability of the diffuse scattering increases with
temperature and saturates at the same value as that of the magnetic
IC modulation.~\cite{Yamada_98}   In either case, the IC diffuse
scattering appears at much higher temperatures than those where the
charge localization behavior has been observed by the NQR wipeout
effect~\cite{Hunt_99} or by conductivity
measurements.~\cite{Komiya_04} 


Our current experiment on the $x=0.05$ sample has established the 
relation between the incommensurate diffuse peak and the superconductivity.  
The previous study by Kimura {\it et al.}~\cite{Kimura_00} on 
samples with $0.10 \leq x \leq 0.12$ can now be interpreted as an 
explicit signature of an incipient lattice modulation of the hole 
doped cuprate superconductors that may be related to the inhomogeneous 
charge and spin state such as the stripes.  
It would be interesting in future studies to examine diffuse 
scattering in $x=0.07$, which is just above the superconducting transition, 
as well as that in optimally doped $x=0.15$ and overdoped $x=0.20$ samples.

\begin{acknowledgments}

The authors thank M. Fujita, G. Gu, A. Kagedan, H. Kimura, C. Stock, 
J. M. Tranquada, and S. Ueki for invaluable discussions.  
The present work was supported by the US-Japan Cooperative Research 
Program on Neutron Scattering.  Work at the
University of Toronto is part of CIAR and supported by NSERC of
Canada, while research at BNL is supported by the U. S. DOE under
contact No. DE-AC02-98CH10886.  The work on SPINS is based upon
activities supported by the National Science Foundation under
Agreement No. DMR-9986442. 

\end{acknowledgments}


\end{document}